\documentclass{PoS}

\newcommand{\SU}{\mathrm{SU}}
\newcommand{\eq}[1]{\begin{equation}\label{#1}}
\newcommand{\en}{\end{equation}}
\newcommand{\ear}[1]{\begin{eqnarray}\label{#1}}
\newcommand{\enar}{\end{eqnarray}}

\title{Recent results in large-$N$ lattice gauge theories}

\ShortTitle{Recent results in large-$N$ lattice gauge theories}

\author{\speaker{Marco Panero}\\
        Department of Physics and Helsinki Institute of Physics, University of Helsinki\\
        FIN-00014 Helsinki, Finland\\
        E-mail: \email{marco.panero@helsinki.fi}
}

\abstract{Generalizations of QCD in which the number of colors $N$ is taken to infinity are characterized by profound mathematical properties, with far-reaching implications for fundamental problems and for phenomenological issues alike. In this contribution, after a brief introduction to the theoretical motivation for studying the large-$N$ limit, the r\^ole of lattice computations in large-$N$ gauge theories is discussed, and a selection of interesting results obtained in recent years is highlighted. Finally, some promising research directions for future studies are pointed out.
\vspace{4cm}
\begin{flushright}
HIP-2012-25/TH
\end{flushright}}

\FullConference{The 30th International Symposium on Lattice Field Theory\\
		 June 24 -- 29,  2012\\
		 Cairns, Australia}

\begin{document}

\section{Introduction and motivation}
\label{sec:intro}

The theoretical investigation of gauge theories in the large-$N$ limit has been a very active research field for almost forty years: the most relevant early works, including the seminal paper by 't~Hooft~\cite{tHooft:1973jz}, are collected in ref.~\cite{Brezin_Wadia}; in addition, several lecture notes and reviews are also available in the literature, see, \emph{e.g.}, refs.~\cite{Makeenko:1999hq, Teper:2009uf}. A new review article---covering, in particular, most of the lattice studies of large-$N$ gauge theories---has recently appeared~\cite{Lucini:2012gg}; the continued interest of the lattice community in this topic is confirmed by the large number of contributions in the parallel sessions of this conference~\cite{GarciaPerez_parallel,
GonzalezArroyo_parallel, Hanada_parallel, Honda_parallel, Kadoh_parallel, Keegan_parallel, Koren_parallel, Lee_parallel, Lohmayer_parallel, Negro_parallel, Okawa_parallel, Orland_parallel}.

\subsection{QCD in the 't~Hooft limit}
\label{subsect:definitions}
Large-$N$ gauge theories are generalizations of QCD in which the number of color charges $N$ is taken to be infinite. In order for this limit to make sense, at least perturbatively, at the same time the coupling $g$ has to be taken to zero, holding the 't~Hooft coupling $\lambda=g^2 N$ fixed. Keeping also the number of quark flavors $n_f$ fixed yields the so-called 't~Hooft limit~\cite{tHooft:1973jz}.\footnote{A different type of large-$N$ limit (Veneziano limit) is obtained, if one also takes the number of quark flavors to infinity, holding $n_f/N$ fixed~\cite{Veneziano:1976wm}. As this limit is generally less easy to study than the 't~Hooft limit, the present discussion will be mostly focused on the latter.}

\begin{figure}[-t]
\centerline{\includegraphics[height=0.10\textheight]{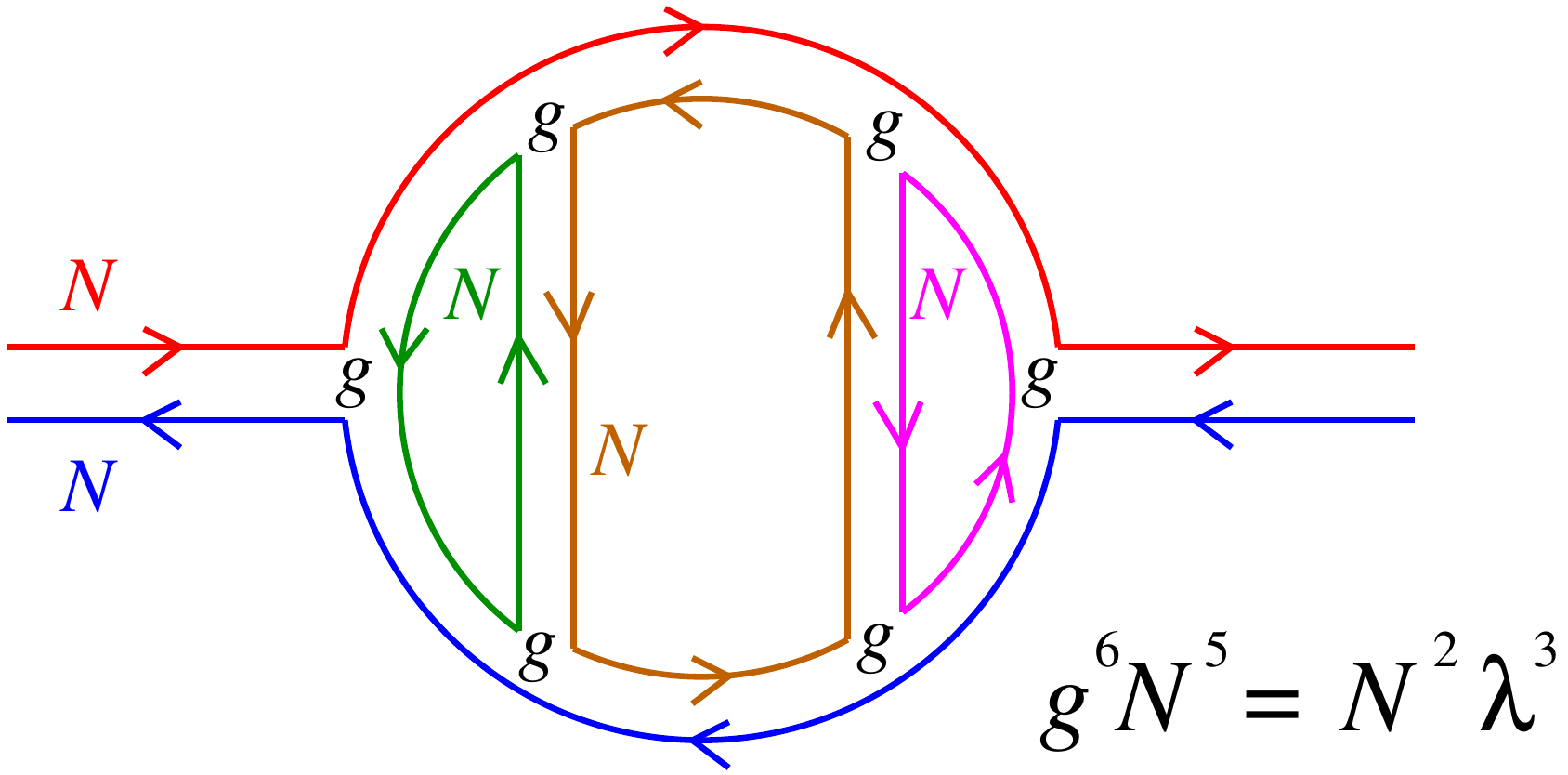} \hfill  \includegraphics[height=0.10\textheight]{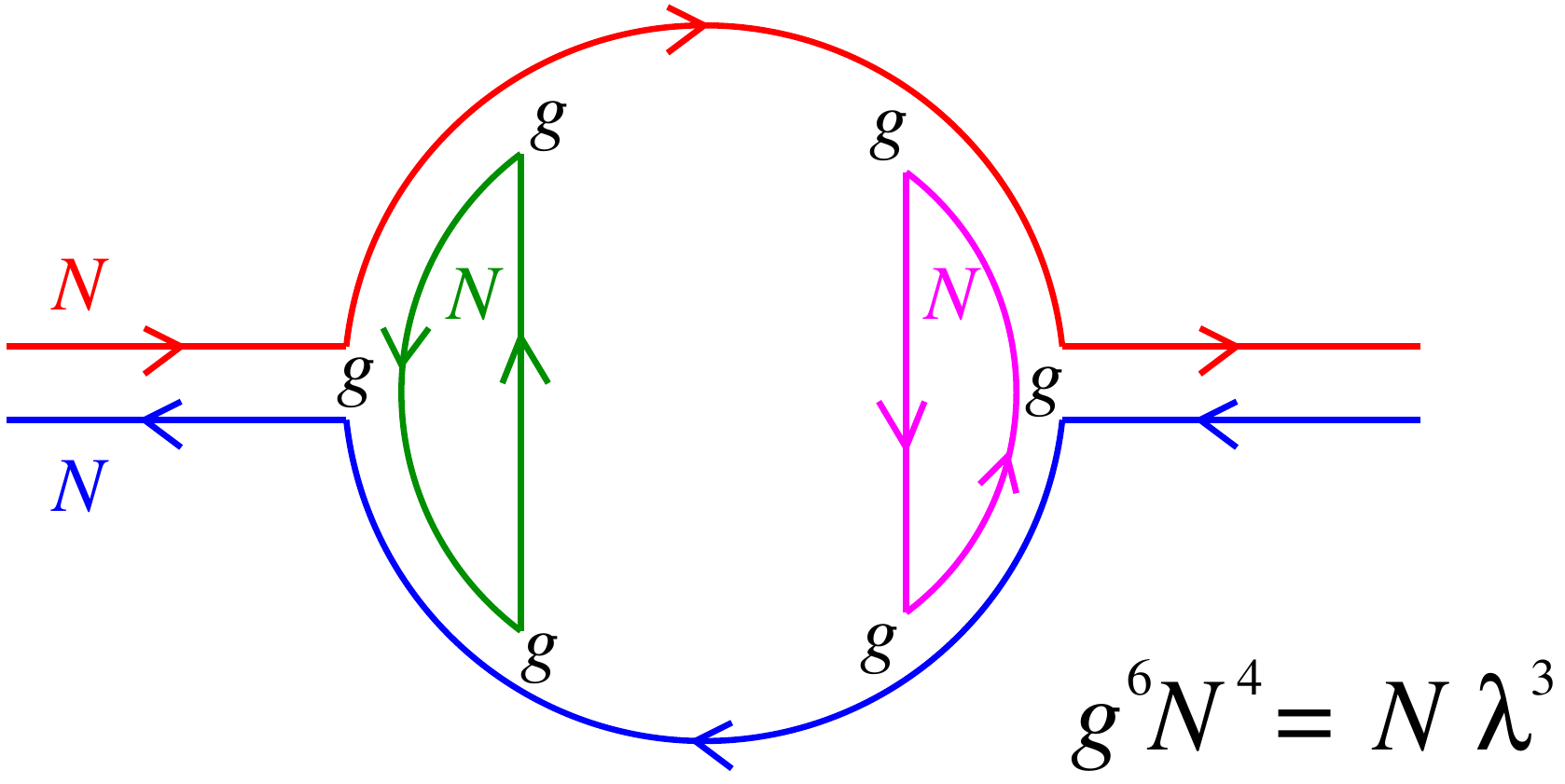} \hfill \includegraphics[height=0.10\textheight]{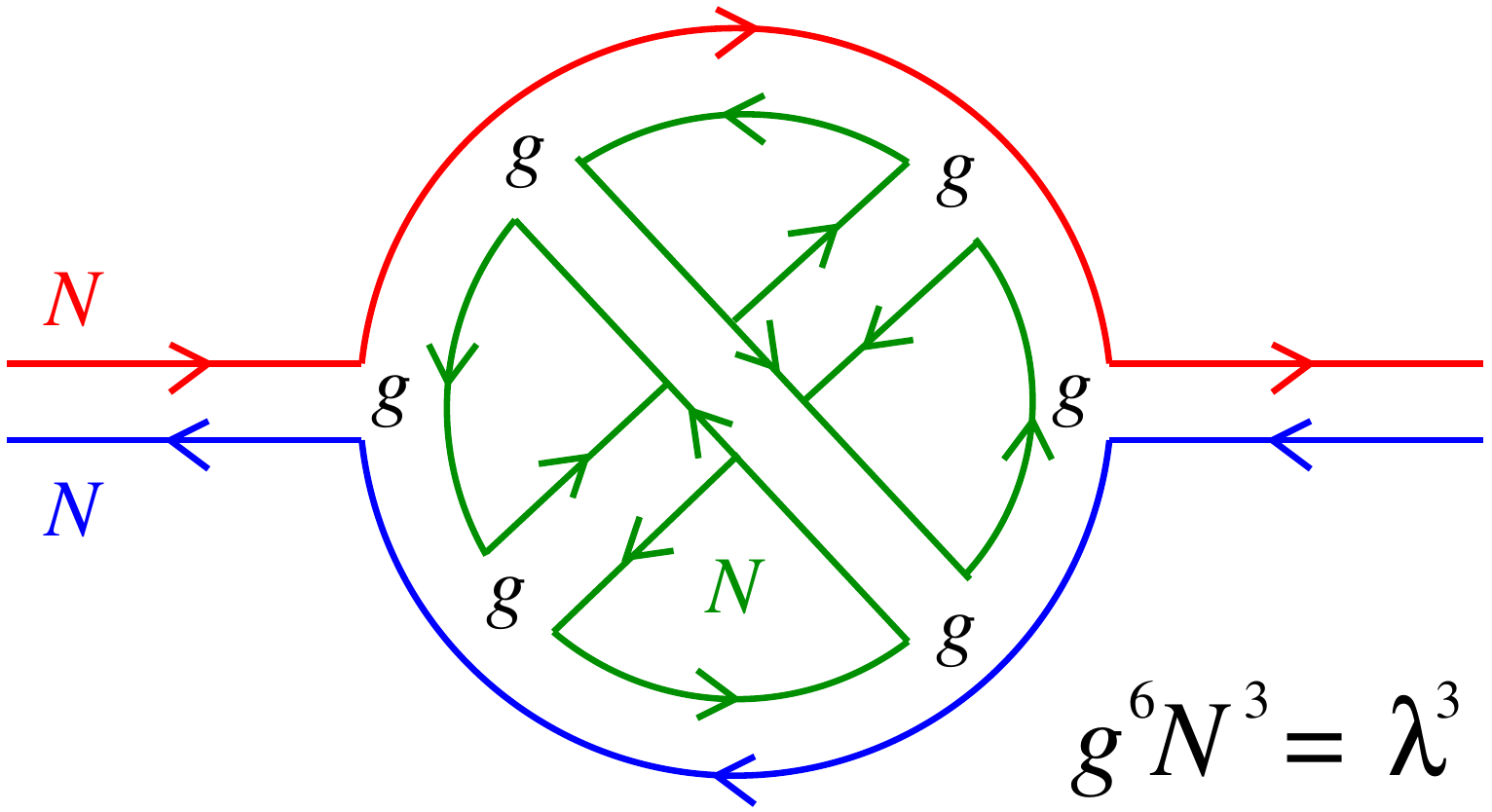}}
\centerline{\includegraphics[height=0.15\textheight]{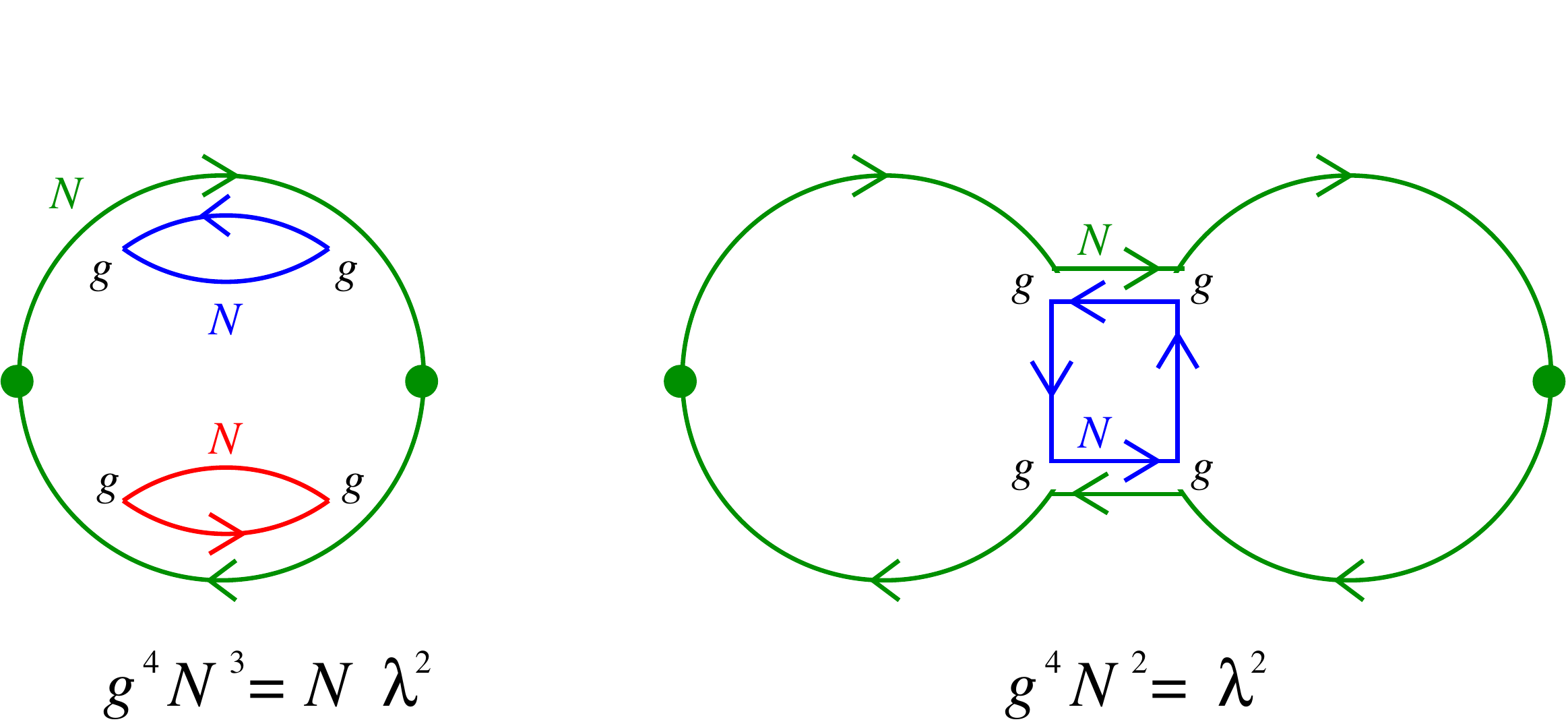}}
\label{fig:planar_quarkloop_nonplanar_and_ozirule}
\caption{Top: Different types of Feynman diagrams contributing to the gluon propagator at three-loop order: in the 't~Hooft limit, the number of independent fundamental color indices is largest for the diagram on the left panel, which only includes planar gluon loops. Replacing a gluon loop with a quark loop (central panel) reduces the power of $N$ by one. Finally, the diagram on the right panel (including only gluon loops, but with a non-planar topology) is $O(1)$. As a consequence, in the 't~Hooft limit the diagrams on the central and right panels are subleading with respect to the one on the left panel. Bottom: Two diagrams contributing to the propagation of a meson (denoted by a blob): the diagram on the right, in which the process goes through an intermediate stage involving only two virtual gluons, is suppressed by a factor of $1/N$ with respect to the one on the left, as expected according to the OZI rule.}
\end{figure}

This double limit is characterized by interesting properties, which dramatically simplify the theory. In particular, the perturbative dynamics is dominated by a special class of Feynman diagrams: counting the powers of $g$ and keeping track of the number of independent fundamental color indices appearing in a Feynman diagram, through the \emph{double-line notation}---which represents propagators of quarks (gluons) in the fundamental (adjoint) representation of the gauge group by single (double) lines---, it is easy to see that the contributions with the largest power of $N$ come from \emph{planar diagrams} (\emph{i.e.}, diagrams which, in this double-line notation, can be drawn on a plane without any crossing lines) \emph{without dynamical quark loops}: see fig.~\ref{fig:planar_quarkloop_nonplanar_and_ozirule} for an example.

More generally, amplitudes for physical processes can be expressed in terms of double series, not only in powers of $\lambda$, but also in powers of $1/N$. The latter expansion has a \emph{topological} nature, as it corresponds to an expansion in classes of diagrams which can be drawn on Riemann surfaces of different topology, \emph{i.e.} with a different number of ``handles'' ($h$) and boundaries ($b$):
\begin{equation}
\label{large_N_amplitude_expansion}
\mathcal{A} = \sum_{h,\,b=0}^{\infty} N^{2-2h-b}\sum_{n=0}^{\infty} c_{(h,\,b),\,n} \lambda^{n}.
\end{equation}
Interestingly, a similar topological expansion is also found in string theory, upon replacing $1/N$ with the string coupling $g_s$. This observation led to speculations that string theory could actually be a reformulation of large-$N$ gauge theory already during the 1970's. From a more modern perspective, this correspondence is also expected to hold in the conjectured duality relating gauge theories and string theory~\cite{Maldacena:1997re, Gubser:1998bc, Witten:1998qj}: in the large-$N$ limit, loop effects on the string side become negligible (see also the plenary contribution by Hanada~\cite{Hanada_plenary}).

\subsection{A wealth of phenomenological implications from large-$N$ counting rules}
\label{subsec:pheno} 

The large-$N$ counting rules introduced above have many phenomenologically interesting implications. In particular, if one \emph{assumes} that QCD in the 't~Hooft limit is a confining theory, it is possible to deduce that the spectrum consists of an infinite number of infinitely narrow mesons and glueballs, with masses $O(1)$. Their interactions are suppressed by powers of $1/\sqrt{N}$, so that large-$N$ QCD is a theory of stable, weakly interacting hadrons. Secondly, exotic states (\emph{e.g.}, tetraquarks, molecules, et c.) are absent. Thirdly, the OZI rule~\cite{Okubo:1963fa, Zweig:1981pd, Iizuka:1966fk} is exact---see fig.~\ref{fig:planar_quarkloop_nonplanar_and_ozirule} for an example.

Another implication of the large-$N$ counting rules is that loop effects in the effective chiral Lagrangian are suppressed, so that it can be studied in the tree-level approximation. Moreover, the axial anomaly turns out to be proportional to $1/N$, and hence suppressed in 't~Hooft's limit.

Baryons can be interpreted as the solitons of large-$N$ QCD, with masses~$O(N)$~\cite{Witten:1979kh}; furthermore, by imposing certain consistency conditions related to unitarity~(see refs.~\cite{Manohar:1998xv, Jenkins:1998wy} and references therein), it is possible to derive a systematic expansion in powers of $1/N$ for quantities such as baryon-meson couplings, baryon masses, magnetic moments, et c.

The large-$N$ limit might also have implications for the QCD phase diagram: in particular, McLerran and Pisarski conjectured the existence of a new, exotic state of matter (\emph{quarkyonic matter}) at large densities~\cite{McLerran:2007qj}---see also ref.~\cite{Lottini:2011zp} for a related lattice model.

Finally, the large-$N$ limit also has a number of interesting implications relevant in the high-energy domain (evolution equations, hadronic cross-sections, parton distributions and structure functions, large-$N$ Standard Model, et c.)~\cite{libro}, that will not be discussed further here.

\subsection{Factorization, volume reduction and large-$N$ equivalences}
\label{subsec:factorization}

Besides interesting phenomenological aspects, large-$N$ gauge theories are also characterized by many intriguing properties at the fundamental level. In particular, the large-$N$ counting rules imply that vacuum expectation values (vev's) of products of gauge-invariant operators are dominated by \emph{disconnected contributions}. This immediately leads to the factorization of vev's of physical operators, up to $O(1/N)$ corrections:
\eq{factorization}
\langle \mathcal{O}_1 \mathcal{O}_2 \rangle = \langle \mathcal{O}_1 \rangle \langle \mathcal{O}_2 \rangle + O(1/N), \label{factorization}
\en
which reveals an analogy with the classical limit of a quantum theory (with $1/N$ playing the r\^ole of $\hbar$). In fact, the analogy can be made explicit, by constructing an appropriate set of coherent states, and a ``classical'' Hamiltonian~\cite{Yaffe:1981vf}.

\begin{figure}[-t]
\centerline{\includegraphics[height=0.25\textheight]{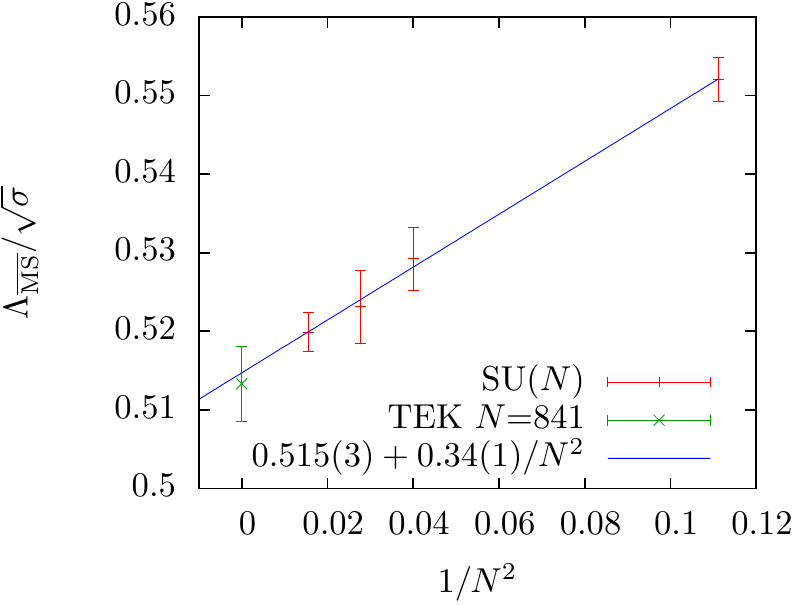} \hfill \includegraphics[height=0.25\textheight]{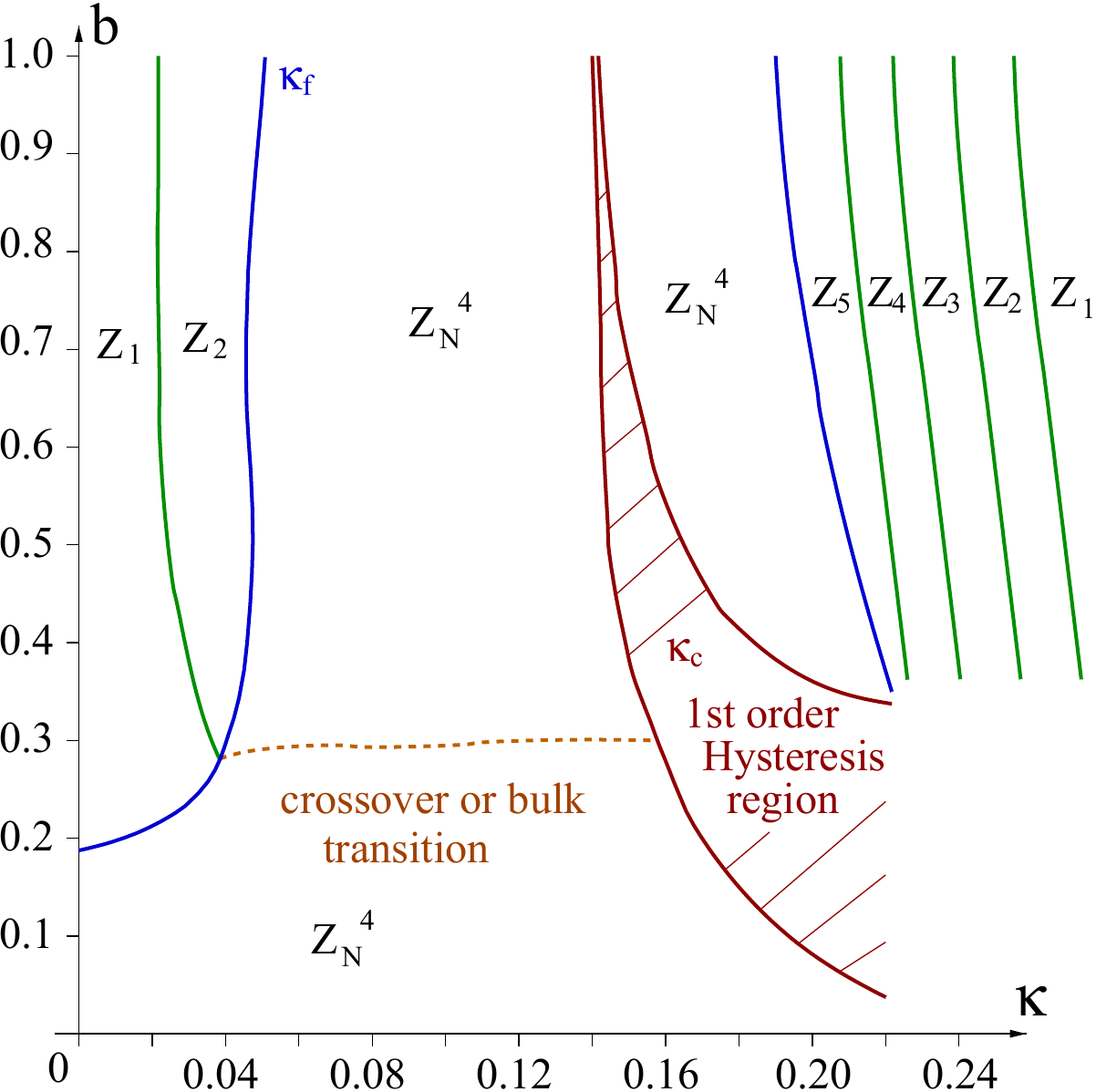}}
\label{fig:TEK_and_funnel}
\caption{Left panel: A comparison of results for the string tension obtained from simulations in a large volume (red symbols) and their linear extrapolation in $1/N^2$ (blue line), with the result for the same quantity from a single-site simulation in the twisted EK model (green symbol), from ref.~\cite{GonzalezArroyo:2012fx}. Right panel: Numerical simulations of large-$N$ gauge theories with dynamical fermions in the adjoint representation of the gauge group in a small volume show evidence that center symmetry is unbroken in a funnel-shaped region that extends to rather large quark masses, and remains of finite width in the continuum limit. The figure, taken from ref.~\cite{Bringoltz:2011by}, is a sketch of the phase diagram based on the results of simulations with $n_f=2$ adjoint flavors.}
\end{figure}

Factorization has an interesting consequence: the Schwinger-Dyson equations satisfied by Wilson loops in the large-$N$ theory on the lattice are independent of the physical hypervolume of the system, \emph{provided center symmetry is unbroken}~\cite{Eguchi:1982nm}: this is the so-called Eguchi-Kawai (EK) volume reduction. In principle, this property would allow one to study the large-$N$ theory in arbitrarily small volumes, either by analytical techniques (reducing the original theory to a matrix model), or by numerical simulations on a single-site lattice. However, it is well-known that center symmetry \emph{does} get broken in a small volume in the continuum limit: this can already be seen at the perturbative level, for all $D>2$. In order to preserve center symmetry, various fixes have been proposed, since the 1980's: for example, in the quenched EK model~\cite{Bhanot:1982sh}, one studies the dynamics of the single-site model for a fixed set of eigenvalues of the link variables along the various directions, and then averages over a center-symmetric distribution for the eigenvalues. However, this method has recently been shown to fail~\cite{Bringoltz:2008av}, due to the fact that the quenching prescription fixes the eigenvalues of the link matrices in the four directions only up to cyclic permutations, and dynamical fluctuations lead to non-trivial correlations among the eigenvalues along different directions. Another approach to preserve center symmetry in the reduced EK model is based on imposing twisted boundary conditions~\cite{GonzalezArroyo:1982hz, GonzalezArroyo:1982ub}: interestingly, this approach can also be used for a non-perturbative definition of field theories in non-commutative spaces~\cite{GonzalezArroyo:1983ac, Aoki:1999vr, Ambjorn:1999ts, Ambjorn:2000nb, Ambjorn:2000cs} (for an alternative regularization of these theories, see, \emph{e.g.}, refs.~\cite{Panero:2006bx, Panero:2006cs} and references therein). Volume independence in the twisted EK model holds both at strong coupling and in the perturbative regime, but, at least for the simplest definition of the twist, it has been found to fail at intermediate couplings, in a range which appears to increase when $N$ grows~\cite{Teper:2006sp, Azeyanagi:2007su}. However, a couple of years ago, the authors who originally proposed the twisted EK model suggested a new formulation of the twist~\cite{GonzalezArroyo:2010ss}, which they are currently studying numerically~\cite{GonzalezArroyo:2012fx}. As an example, the plot on the left panel of fig.~\ref{fig:TEK_and_funnel}, taken from ref.~\cite{GonzalezArroyo:2012fx}, shows how the extrapolation of results for the string tension obtained from simulations in large volume (at moderately large $N$) compares with the result from a single-site simulation in the new version of the twisted model, at a much larger value of $N$: the results appear to be in nice agreement, and motivate further studies of this model.

Another possibility to preserve center symmetry in EK models is based on the inclusion of dynamical adjoint fermions (with periodic boundary conditions in all directions): this idea was initially proposed in ref.~\cite{Kovtun:2007py}, and has since been studied both analytically and numerically by a number of authors~\cite{Lee_parallel, Okawa_parallel, Hollowood:2009sy, Azeyanagi:2010ne, Catterall:2010gx, Bringoltz:2009kb, Bringoltz:2011by, Cossu:2009sq, Hietanen:2009ex, Hietanen:2010fx, Armoni:2011dw}. Recent results indicate that, indeed, EK volume reduction with adjoint Dirac fermions works as expected, both with $n_f=1$ and $n_f=2$ flavors: see, \emph{e.g.}, the plot on the right panel of fig.~\ref{fig:TEK_and_funnel}, taken from ref.~\cite{Bringoltz:2011by}, which shows the center symmetry realizations in the $n_f=2$ model, as a function of the hopping parameter $\kappa$ and of the bare lattice coupling $b=1/\lambda$.

Another way to enforce center symmetry in EK models is based on double-trace deformations~\cite{Unsal:2008ch}: one modifies the usual Yang-Mills (YM) action adding (products of) traces of Polyakov loops, with positive coefficients. This suppresses center-symmetry breaking configurations, at the cost of $O(1/N^2)$ corrections to the observables:
\begin{equation}
S_{\mbox{\tiny{YM}}} \longrightarrow S_{\mbox{\tiny{YM}}} + \frac{1}{N_t^3} \sum_{\vec{x}} \sum_{n=1}^{\lfloor N/2 \rfloor} a_n \left| \mbox{tr}(L^n(\vec{x})) \right|^2.
\end{equation}
Related ideas have also been discussed in the context of $\SU(N)$ YM theory at finite temperature~\cite{Myers:2007vc, Ogilvie_parallel}. Dedicated numerical algorithms to study the EK model with double-trace deformation have been devised~\cite{Vairinhos:2010ha}, and preliminary investigations are currently under way~\cite{volume_reduction_paper}. A nice feature of this approach is that one can reduce only one (or a few) direction(s), while keeping the others large.

Finally, Neuberger and collaborators proposed the partial reduction approach to EK~\cite{Narayanan:2003fc, Kiskis:2003rd}: the idea is to simulate the large-$N$ theory in lattices which are small, but still larger than the critical size corresponding to the inverse of the deconfinement temperature $T_c$. As an example of results obtained in this approach, in ref.~\cite{Kiskis:2009rf} the confining potential was computed up to distances equal to $9$ lattice spacings, from simulations on a lattice of linear size $L=6a$.

Volume reduction and volume independence in large-$N$ gauge theories can be interpreted as an example of ``orbifold'' equivalence~\cite{Bershadsky:1998cb, Strassler:2001fs}, namely as a correspondence based on projections under some discrete subgroup of the global symmetries of two different theories~\cite{Kovtun:2007py, Neuberger:2002bk, Kovtun:2003hr, Kovtun:2004bz, Kovtun:2005kh, Unsal:2006pj}: under the assumption that the discrete symmetry used in this projection is not spontaneously broken, the vev's and correlation functions of invariant (or ``neutral'') sectors of observables in the original (``parent'') and projected (``daughter'') theories are equal---up to a trivial rescaling of coupling constants and volume factors. Such orbifold equivalences do not relate only theories defined in different volumes, but also theories with different field content: for example, the orientifold planar equivalence~\cite{Armoni:2003gp, Armoni:2003fb} (investigated on the lattice in ref.~\cite{Lucini:2010kj}) can be interpreted as a correspondence between two different daughter theories obtained by orbifold projections from a common parent theory~\cite{Unsal:2006pj}. Finally, orbifold projections are also relevant for lattice formulations of supersymmetry~\cite{Catterall:2009it} (see also refs.~\cite{Hanada:2007ti, Anagnostopoulos:2007fw, Ishii:2008ib, Ishiki:2008te} for work on related topics).

\section{A selection of recent results}
\label{sec:results}

In this section, we discuss a selection of recent lattice results obtained from simulations of large-$N$ gauge theories (in a large volume).

\subsection{Results in four spacetime dimensions}
\label{subsec:4D}

\begin{figure}[-t]
\centerline{\includegraphics[height=0.17\textheight]{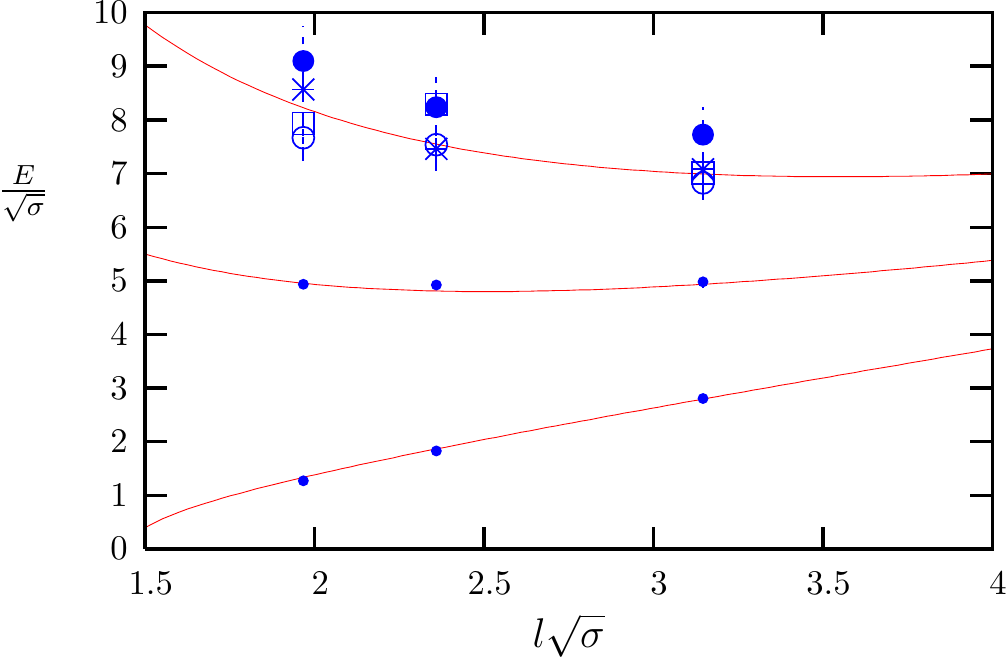} \includegraphics[height=0.17\textheight]{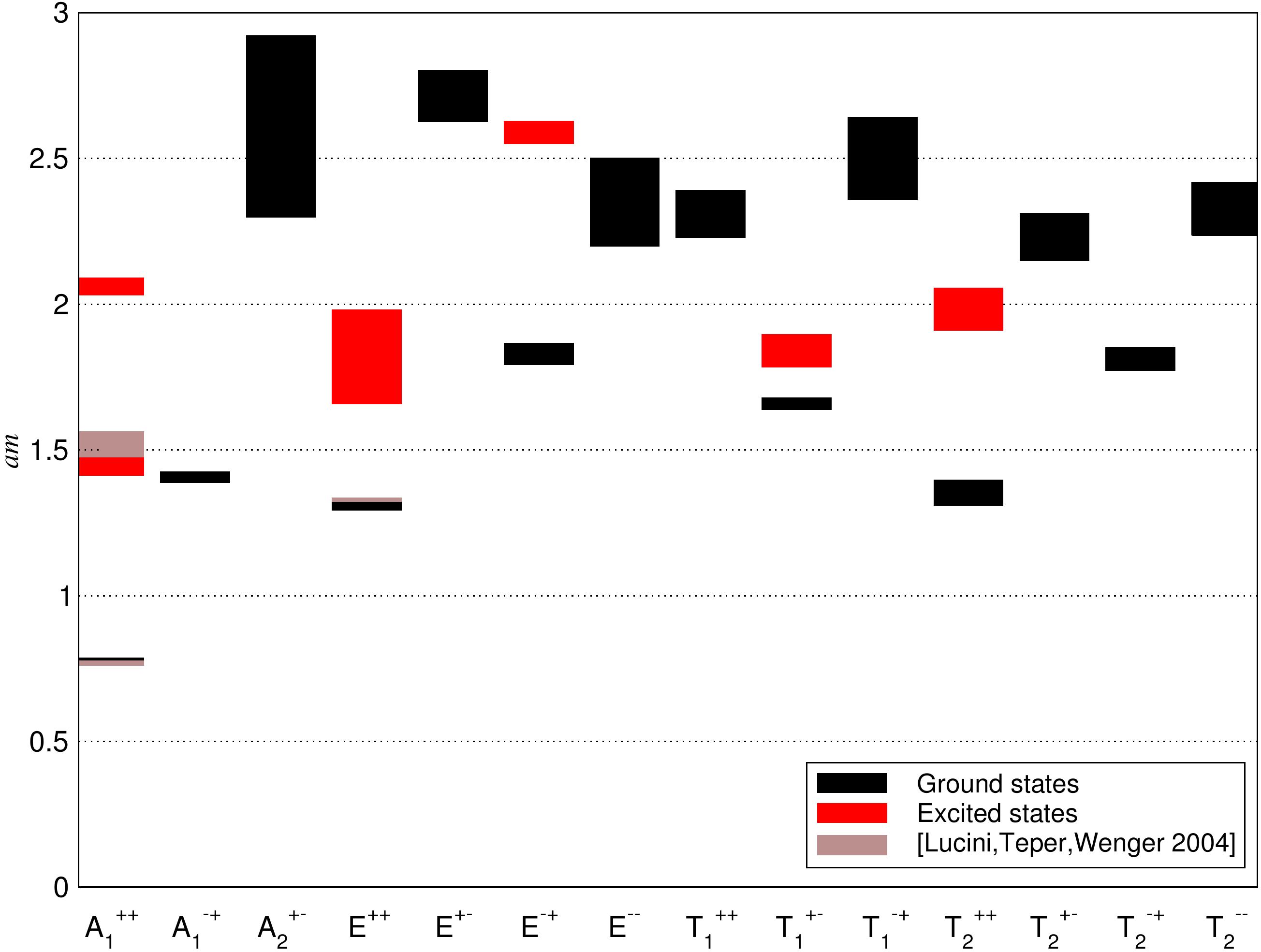}  \includegraphics[height=0.17\textheight]{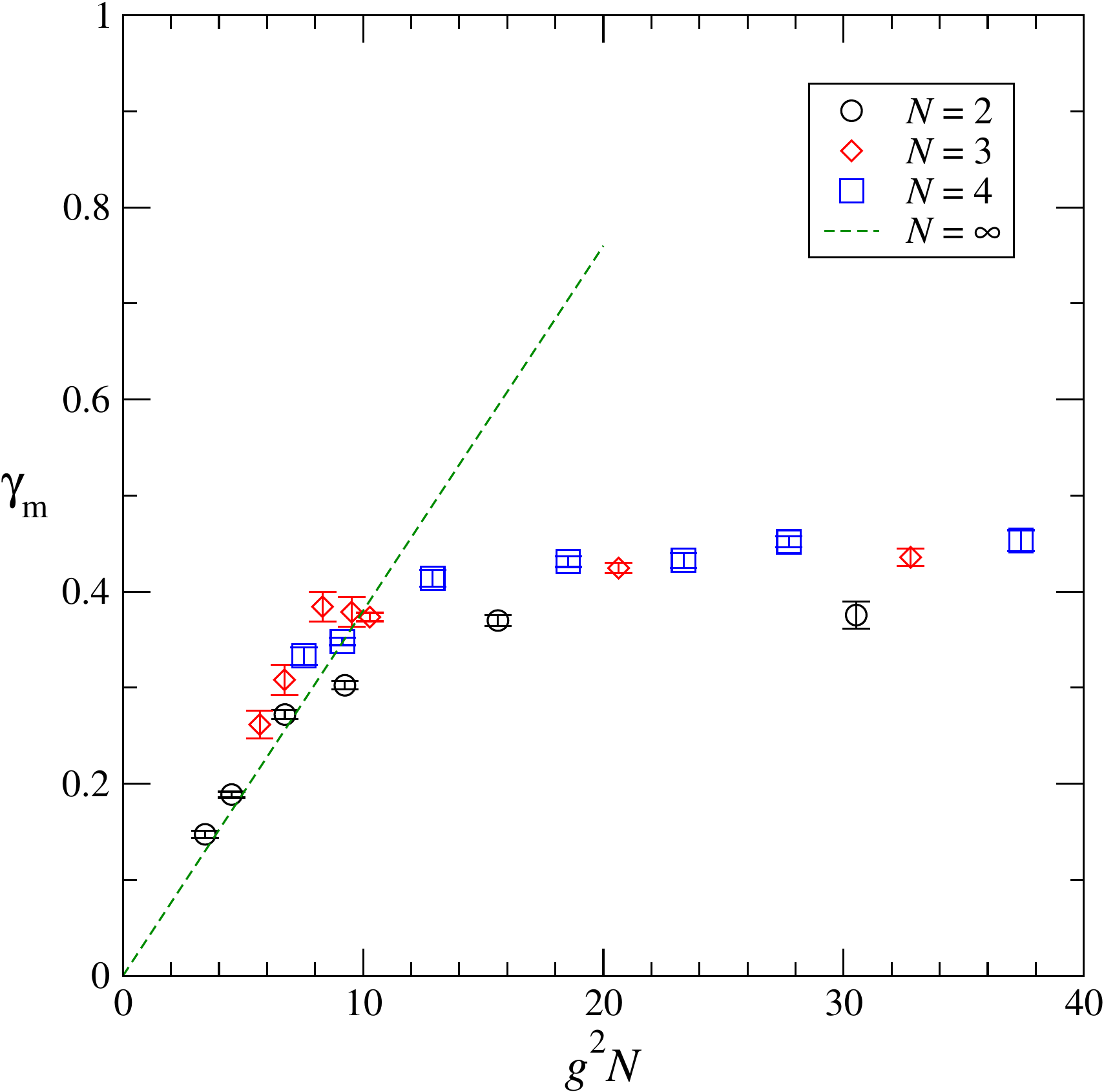}}
\caption{Left panel: Torelon spectrum in $\SU(5)$ YM theory, taken from ref.~\cite{Athenodorou:2010cs}, in comparison with the predictions from the Nambu-Goto effective string model. Central panel: Large-$N$ YM glueball spectrum, taken from ref.~\cite{Lucini:2010nv}; the figure also shows the results for the ground state and first excitation in the $J^{PC}=0^{++}$ channel, and for the $2^{++}$ ground state taken from ref.~\cite{Lucini:2004my}. Right panel: Dependence of the mass anomalous dimension in QCD$_N$, with different numbers of colors and $2$ flavors of dynamical fermions in the two-index symmetric representation of the gauge group, on the renormalized 't~Hooft coupling in the SF scheme, from ref.~\cite{DeGrand:2012qa}. The dashed green line denotes the leading-order perturbative prediction in the large-$N$ limit.}
\label{fig:4D_strings_glueballs_and_Svetitsky}
\end{figure}

As mentioned in subsec.~\ref{subsec:pheno}, many phenomenologically interesting implications of the large-$N$ counting rules are derived under the assumption that QCD (or YM) is confining in the large-$N$ limit. Testing the correctness of this assumption at the non-perturbative level, via lattice simulations, was one of the first problems to be addressed, and by now we know that, indeed, $\SU(N)$ gauge theories are confining in the 't~Hooft limit~(see, \emph{e.g.}, refs.~\cite{DelDebbio:2001sj, Meyer:2004hv}). Lattice studies also show that confining flux tubes can be modelled quite accurately as Nambu-Goto strings (a fact generically observed in confining theories~\cite{Bali:1994de, Luscher:2002qv, Juge:2002br, Koma:2003gi, Panero:2004zq, Panero:2005iu, Bonati:2011nt, Amado:2012wt}). As an example, the plot on the left panel of fig.~\ref{fig:4D_strings_glueballs_and_Svetitsky}, taken from ref.~\cite{Athenodorou:2010cs}, shows a comparison of the torelon spectrum in $\SU(5)$ YM theory with the predictions from the Nambu-Goto model. Other works studying the effective string picture at large $N$ include those by Lucini and Teper~\cite{Lucini:2001nv}, by Lohmayer and Neuberger~\cite{Lohmayer:2012ue} and by Mykk\"anen~\cite{Mykkanen:2012dv}.

As for the YM hadron spectrum, glueball masses turn out to have a smooth dependence on $N$. The central panel of fig.~\ref{fig:4D_strings_glueballs_and_Svetitsky}, taken from ref.~\cite{Lucini:2010nv}, shows the results extrapolated to the 't~Hooft limit for the masses of ground-state and excited glueballs, in several different channels. The results from an earlier study~\cite{Lucini:2004my} are also shown.

Another interesting issue deserving non-perturbative investigation is the dependence of the large-$N$ theory on the 't~Hooft coupling. As mentioned in subsec.~\ref{subsect:definitions}, the expectation that $\lambda$ is the appropriate coupling for the 't~Hooft limit of the theory is based on perturbative arguments. Hence, one may wonder, whether this is borne out non-perturbatively. Lattice simulations do confirm that this is the case. This can already be seen from the bare lattice coupling: ref.~\cite{Allton:2008ty} studied how the tree-level improved bare lattice 't~Hooft coupling runs with the lattice spacing $a$ (determined non-perturbatively from computations of the string tension) in YM theories with a different number of colors: the collapse of data obtained from simulations at different values of $N$ showed clearly that $\lambda$ is an appropriate coupling for the large-$N$ theory. The running coupling in the Schr\"odinger functional (SF) scheme~\cite{Luscher:1991wu, Luscher:1992an} in $\SU(4)$ YM theory was studied in ref.~\cite{Lucini:2008vi}, which found that, after converting to the $\overline{\mbox{MS}}$ scheme, the renormalized coupling nicely approaches the two-loop perturbative prediction as the momentum scale, at which it is defined, is increased.

Motivated by the recent interest in walking technicolor theories~\cite{DelDebbio:2010zz, Rummukainen:2011xv, Giedt_plenary}, the authors of ref.~\cite{DeGrand:2012qa} studied the mass anomalous dimension $\gamma_m$ in generalizations of QCD based on gauge group $\SU(N)$ (with $N=2$, $3$ and $4$) with two dynamical flavors of fermions in the two-index symmetric representation. Their results, displayed in the right panel of fig.~\ref{fig:4D_strings_glueballs_and_Svetitsky}, reveal clear similarities among the theories with a different number of colors.

\begin{figure}[-t]
\centerline{\includegraphics[height=0.22\textheight]{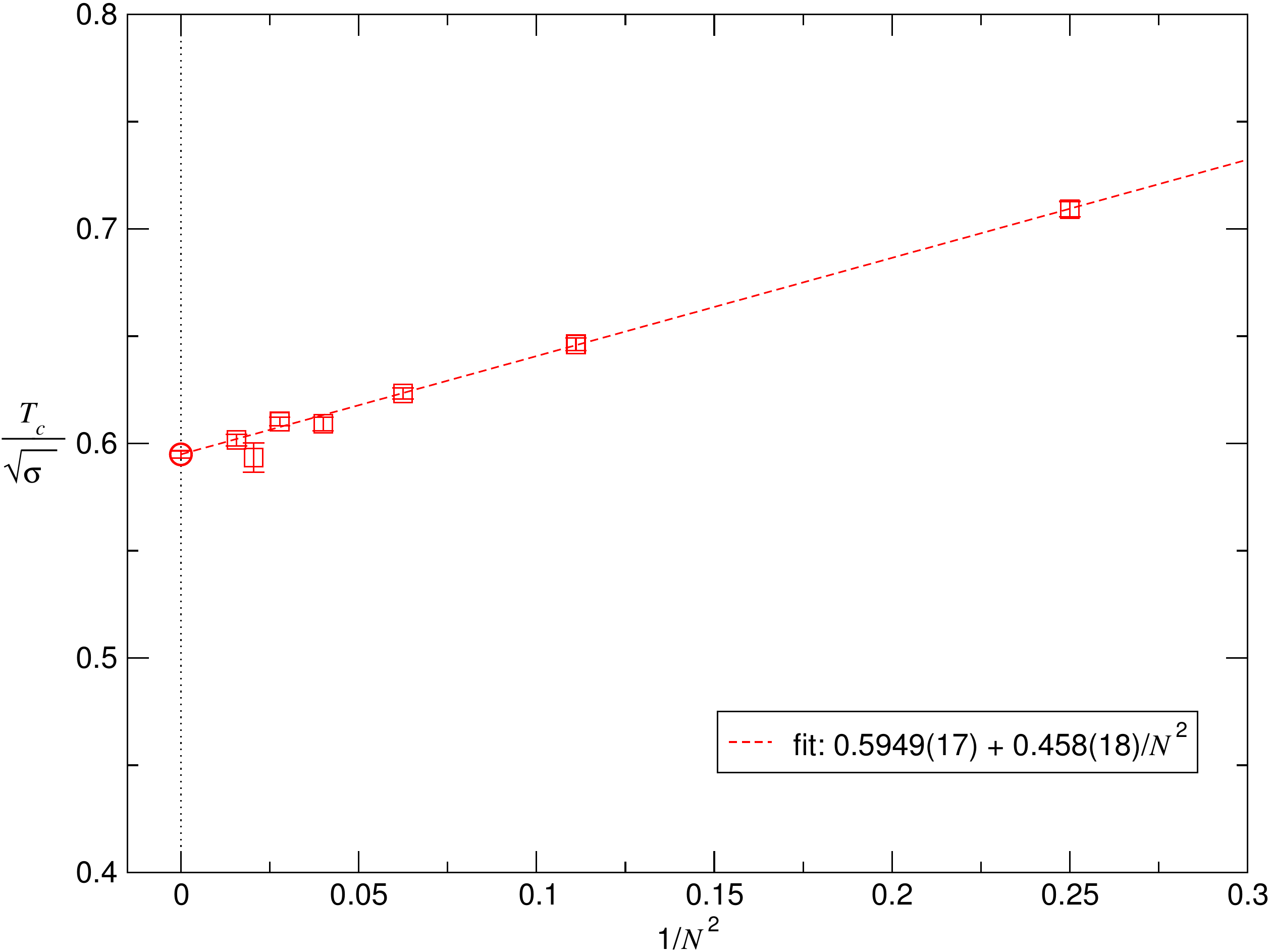} \hfill \includegraphics[height=0.22\textheight]{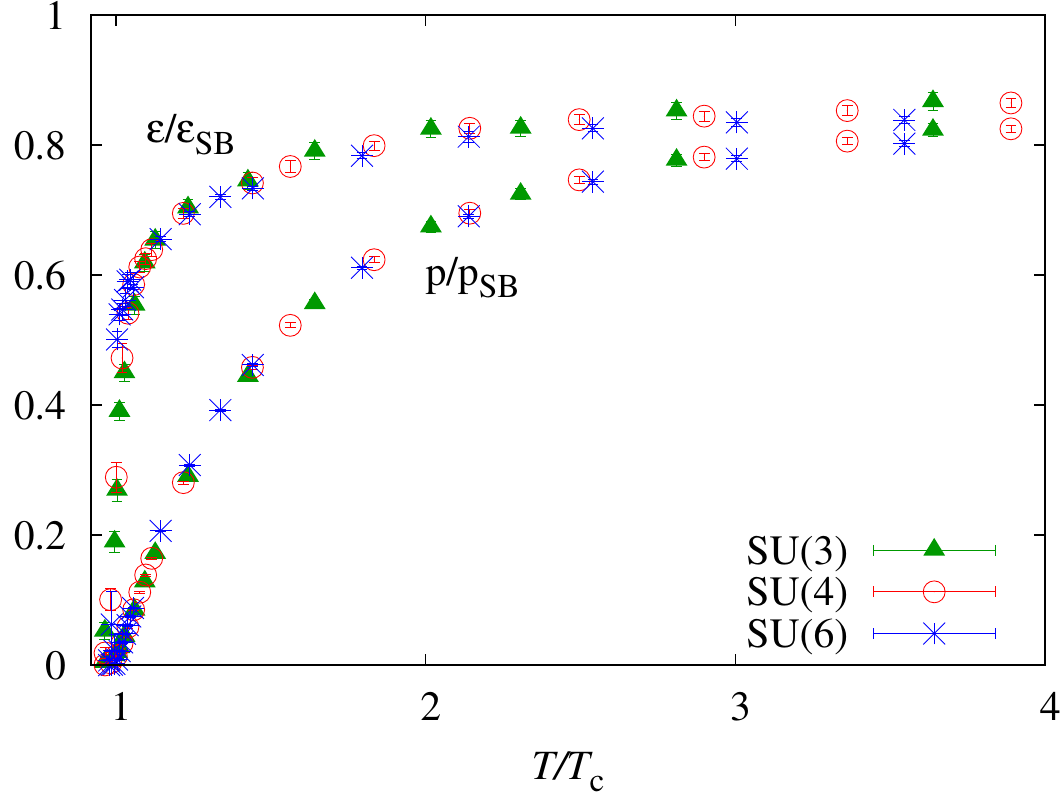}}
\caption{Left panel: Extrapolation of the ratio of the critical deconfinement temperature over the square root of the string tension to the large-$N$ limit, from ref.~\cite{Lucini:2012wq}. Right panel: The pressure ($p$) and energy density ($\epsilon$), normalized to their respective SB limits, as a function of $T/T_c$, in YM theories with $N=3$, $4$ and $6$ colors, from ref.~\cite{Datta:2010sq}.}
\label{fig:Tc_over_root_sigma_and_Datta_Gupta}
\end{figure}

Lattice simulations of large-$N$ Yang-Mills theories at finite temperature $T$ have been carried out in several works~\cite{Lucini:2002ku, Lucini:2003zr, Lucini:2005vg, Bursa:2005yv, Bringoltz:2005rr, Bringoltz:2005xx, Panero:2008mg, Panero:2009tv, Datta:2009jn, Datta:2010sq, Mykkanen:2012ri, Lucini:2012wq}: they revealed that all these theories have a finite-temperature deconfinement transition at a critical temperature $T_c$, which has a smooth dependence on $N$ (see the plot on the left panel of fig.~\ref{fig:Tc_over_root_sigma_and_Datta_Gupta}, taken from ref.~\cite{Lucini:2012wq}). The transition is of first order for $N \ge 3$, with a latent heat $O(N^2)$ in the large-$N$ limit. 

The equation of state has been investigated in various studies~\cite{Bringoltz:2005rr, Panero:2009tv, Datta:2010sq}: in the deconfined phase, the main equilibrium thermodynamic quantities, when normalized to the Stefan-Boltzmann limit (which includes a $N^2-1$ factor, related to the number of gluon degrees of freedom in the free theory) and plotted as a function of $T/T_c$, have almost no (residual) dependence on the number of colors $N$: see the plot on the right panel of fig.~\ref{fig:Tc_over_root_sigma_and_Datta_Gupta}, taken from ref.~\cite{Datta:2010sq}, as an example. 

Similarly, good scaling properties with the number of colors hold for renormalized Polyakov loops in the deconfined phase, and their free energies for different representations are in excellent agreement with Casimir scaling~\cite{Ambjorn:1984mb}, as recently shown by Mykk\"anen \emph{et al.} in ref.~\cite{Mykkanen:2012ri}. 

The topological properties and the $\theta$-dependence of large-$N$ gauge theories have been studied in various works~\cite{Lucini:2001rc, Cundy:2002hv, DelDebbio:2002xa, Lucini:2004yh}---see also the review~\cite{Vicari:2008jw}. The topological susceptibility has a non-vanishing value in the 't~Hooft limit, with small $O(1/N^2)$ corrections at finite $N$.

\begin{figure}[-t]
\centerline{\includegraphics[width=0.4\textwidth]{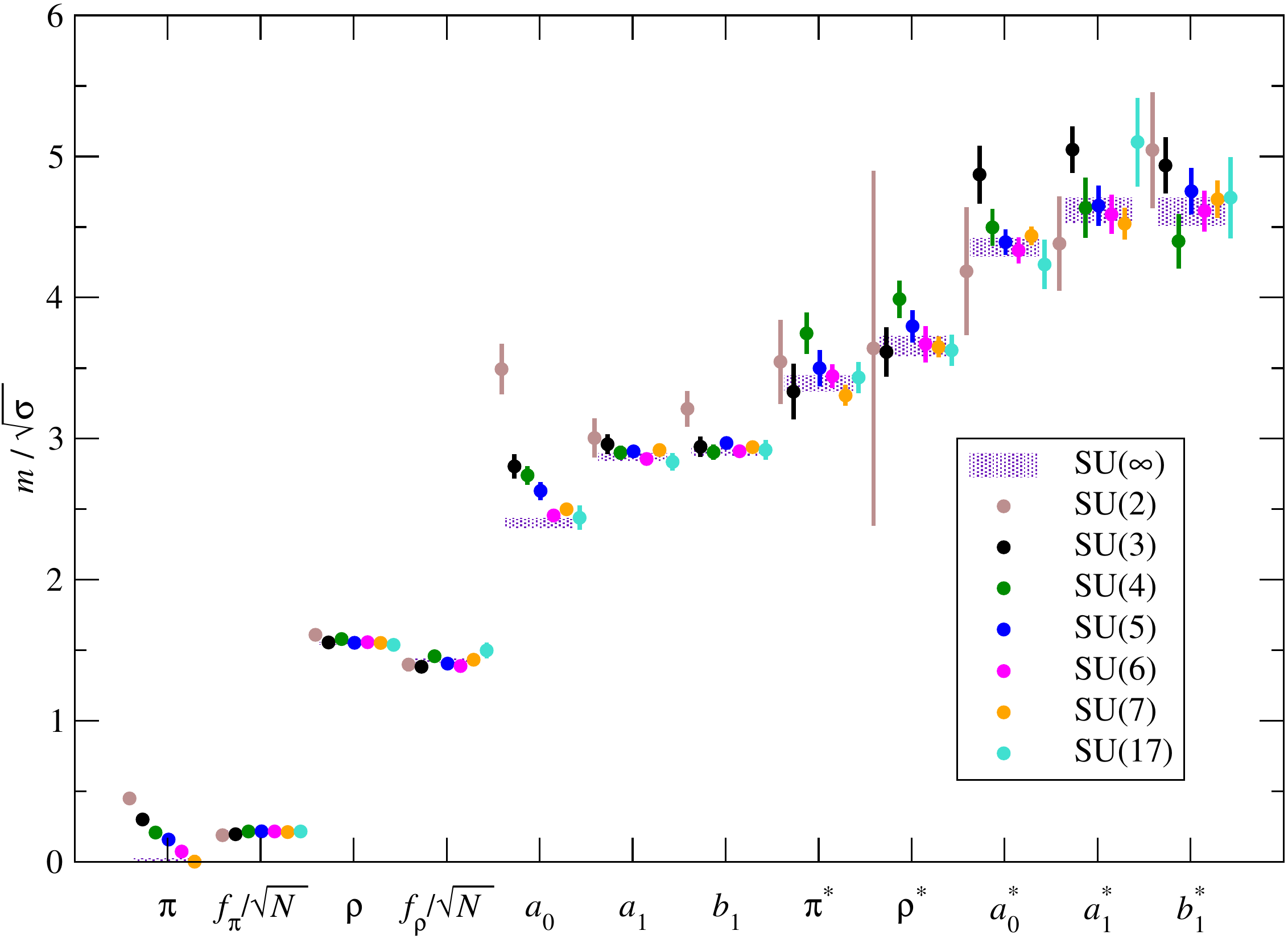} \hfill \includegraphics[width=0.55\textwidth]{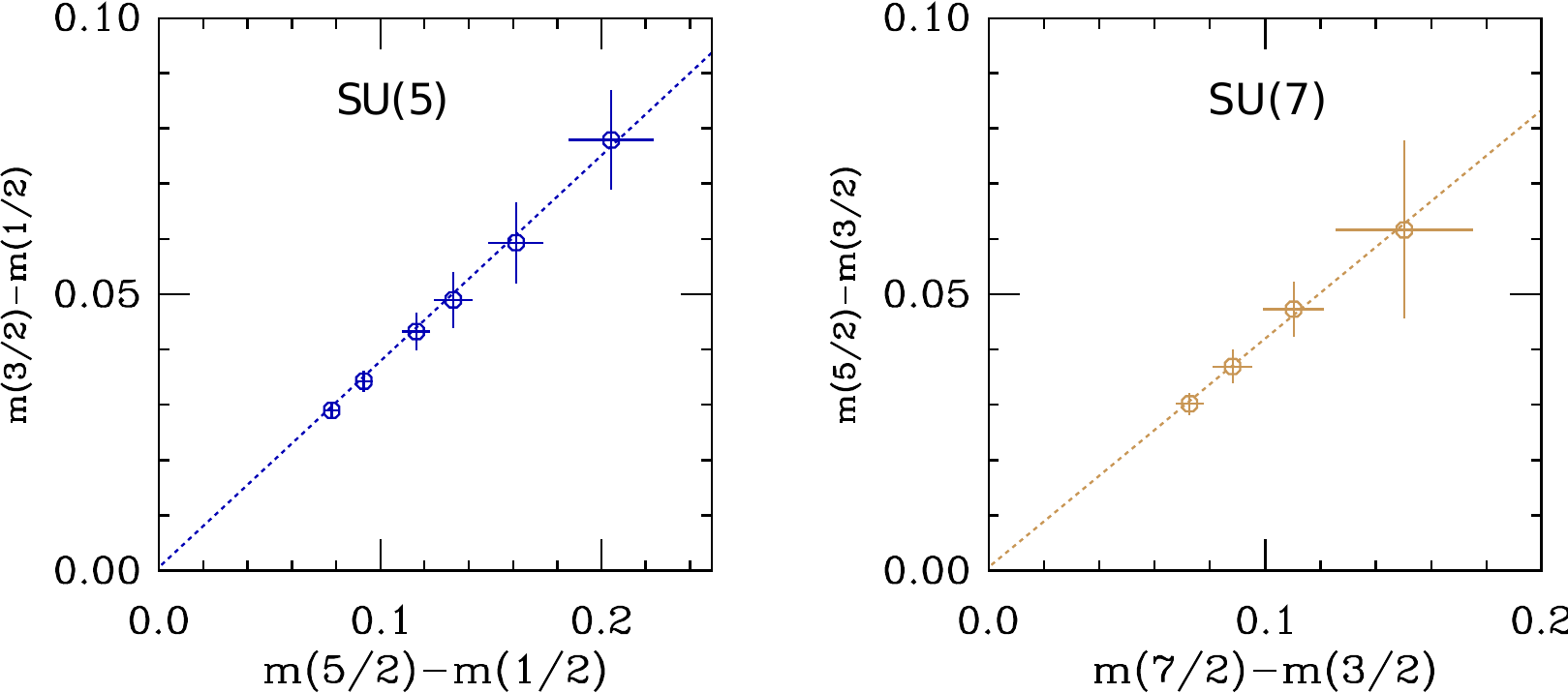}}
\caption{Left panel: Quenched mesonic spectrum at large $N$, taken from ref.~\cite{Bali:2013kia}. Right panel: Evidence for a rotor spectrum in large-$N$ baryons of different spin, from quenched computations in QCD with $N=5$ and $N=7$ colors~\cite{DeGrand:2012hd}.}
\label{fig:Luca_and_DeGrand}
\end{figure}

Finally, there exist quenched computations of the spectrum of large-$N$ mesons~\cite{DelDebbio:2007wk, Bali:2008an, Hietanen:2009tu, Bali:2013kia} (see, \emph{e.g.}, the plot on the left panel of fig.~\ref{fig:Luca_and_DeGrand}, taken from ref.~\cite{Bali:2013kia}) and baryons: in particular, the plot on the right panel of fig.~\ref{fig:Luca_and_DeGrand}, from ref.~\cite{DeGrand:2012hd}, shows clear evidence for a rotor spectrum (as expected from the arguments in refs.~\cite{Adkins:1983ya, Jenkins:1993zu}) in baryons of different spin in theories with $N=5$ and $N=7$ colors. In addition, some analytical large-$N$ predictions for baryons~\cite{Jenkins:1998wy} were compared with results from unquenched $N=3$ QCD simulations in refs.~\cite{Jenkins:2009wv, WalkerLoud:2011ab}.

\subsection{Results in three spacetime dimensions}
\label{subsec:3D}

\begin{figure}[-t]
\centerline{\includegraphics[height=0.25\textheight]{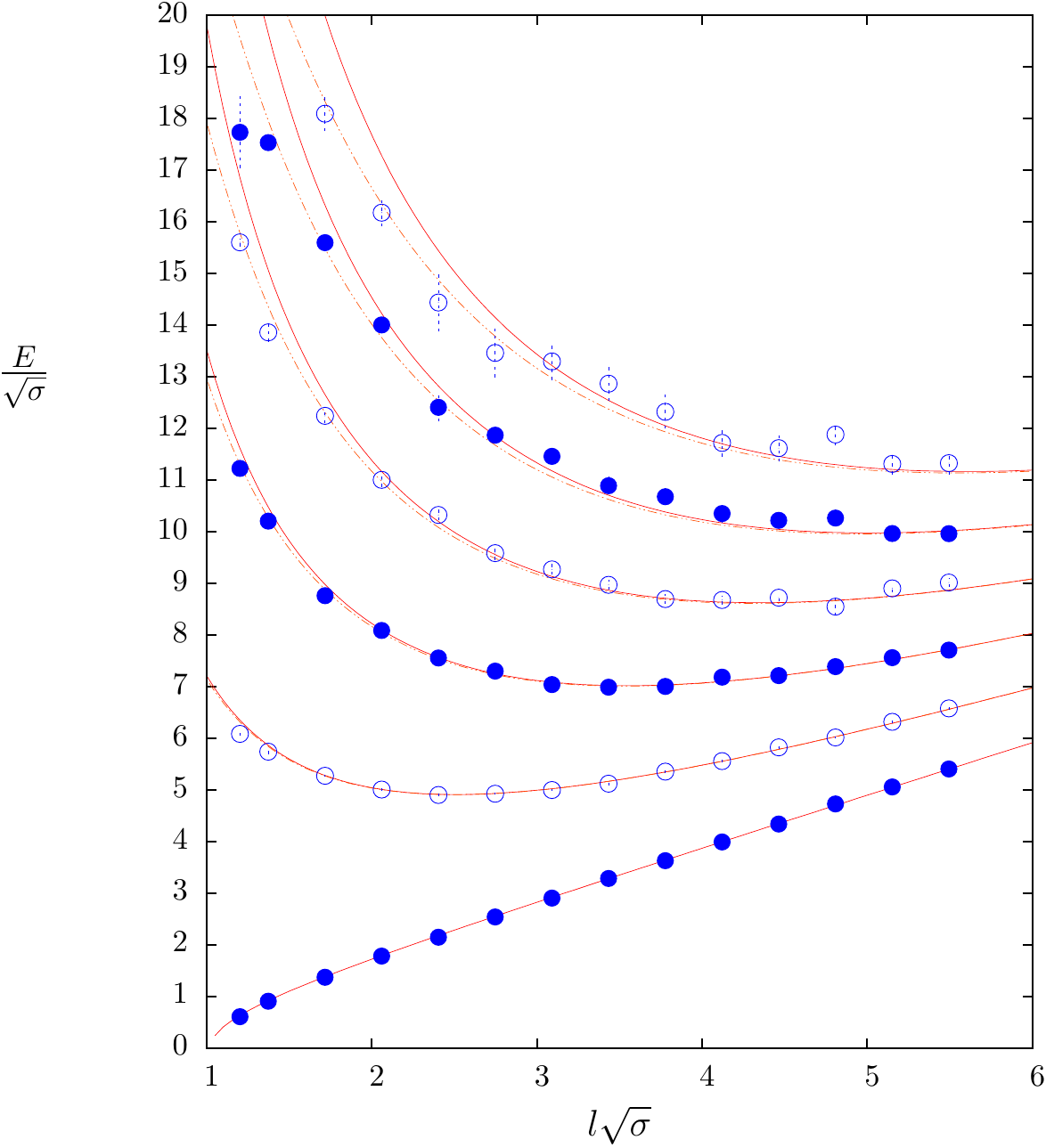} \hfill \includegraphics[height=0.25\textheight]{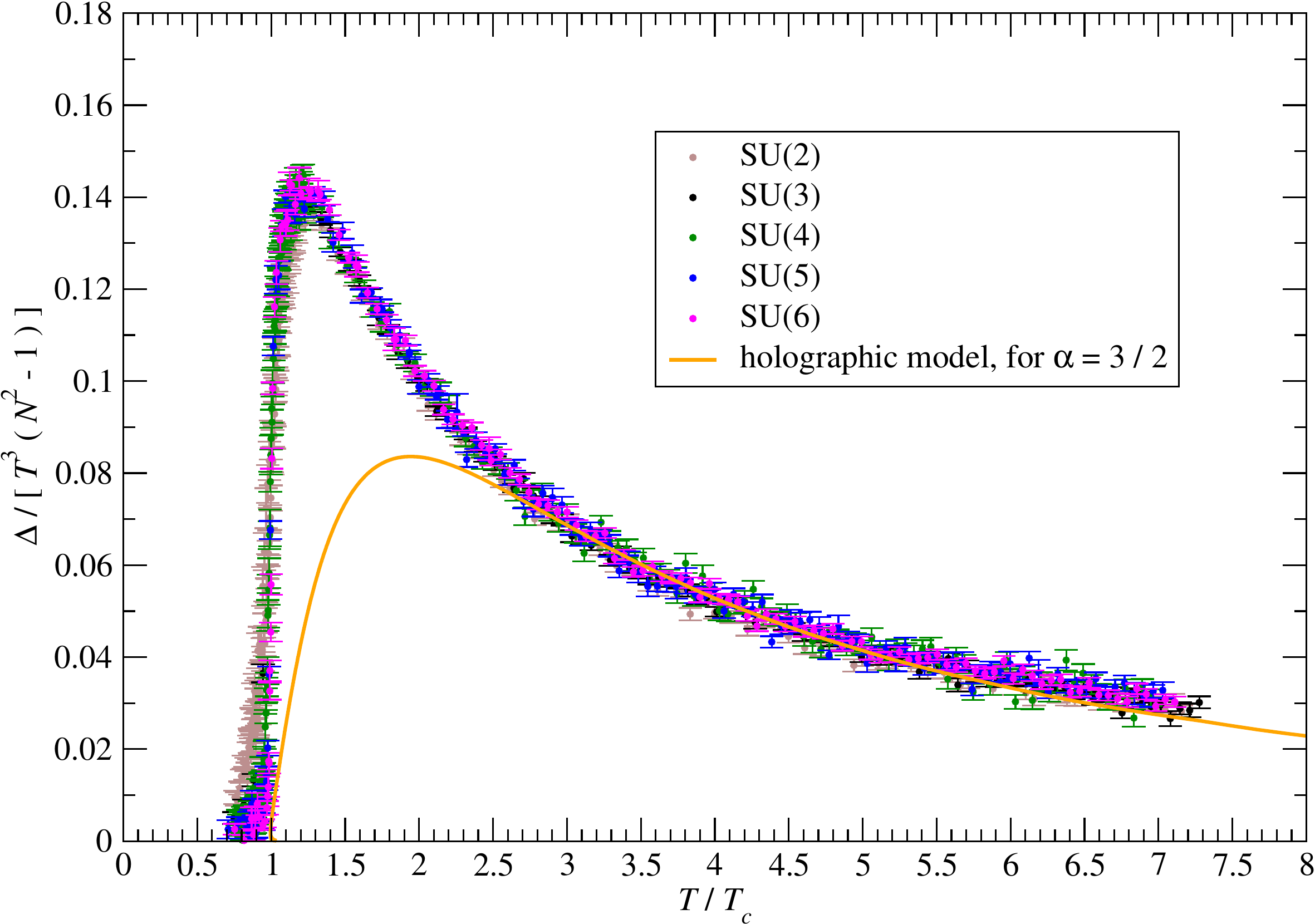}}
\caption{Results in 3D YM theories: The left panel, from ref.~\cite{Athenodorou:2011rx}, shows a comparison of the spectrum of confining flux tubes in $\SU(6)$ YM theory with the predictions of the Nambu-Goto effective string model. The right panel, taken from ref.~\cite{Caselle:2011mn}, displays the trace of the energy-momentum tensor per gluon (in units of $T^3$) as a function of $T/T_c$. The solid yellow line is the prediction from a 3D holographic model inspired by the improved holographic QCD model~\cite{Gursoy:2007cb, Gursoy:2007er, Gursoy:2008bu} (see also ref.~\cite{Alanen:2009xs}).}
\label{fig:3D_strings_and_EoS}
\end{figure}

Large-$N$ gauge theories in three spacetime dimensions (3D) share many qualitative features with their four-dimensional counterparts. In particular, the 't~Hooft limit of these theories has been shown to be confining since many years~\cite{Teper:1998te}, and, similarly to what happens in four dimensions, confining flux tubes can be described quite accurately in terms of Nambu-Goto strings, as shown in the left panel of fig.~\ref{fig:3D_strings_and_EoS}, taken from ref.~\cite{Athenodorou:2011rx}. Other recent studies of string effects in large-$N$ gauge theories in three dimensions include works by Caselle \emph{et al.}~\cite{Caselle:2011vk} and by Mykk\"anen~\cite{Mykkanen:2012dv}. 

In fact, string-like behavior has been observed in a variety of confining 3D models~\cite{Caselle:1996ii, Caselle:2002ah, Majumdar:2002mr, Juge:2004xr, Caselle:2004er, Gliozzi:2005ny, Giudice:2007sk, Giudice:2009di, Brandt:2009tc, Brandt:2010bw, Rajantie:2012zn} (see ref.~\cite{Kuti:2005xg} for a review): these studies confirm that the low-energy dynamics of confining flux tubes is consistent with the Nambu-Goto model at the lowest orders in an expansion around the long-string limit~\cite{Luscher:1980fr, Luscher:1980ac, Luscher:1980iy, Polchinski:1991ax, Luscher:2004ib, Drummond:2004yp}, and are currently reaching levels of precision sufficient to reveal the deviations expected at high orders~\cite{Aharony:2010cx, Aharony:2010db, Aharony:2011gb, Billo:2012da, Gliozzi:2012cx}.

For the YM spectrum, there is large numerical evidence that, like in four dimensions, glueball masses have smooth, finite large-$N$ limits also in 3D~\cite{Teper:1998te, Meyer:2004gx}. A very mild dependence on $N$ has also been observed for gluon and ghost propagators in Landau gauge~\cite{Maas:2010qw}. 

Finally (and, again, similarly to the four-dimensional case), also the 3D equation of state appears to have only a trivial dependence on the number of colors~\cite{Caselle:2011fy, Caselle:2010qd, Caselle:2011mn}, as shown, for example, by the plot of the trace of the energy-momentum tensor per gluon d.o.f. and in units of $T^3$, as a function of $T/T_c$, on the right panel of fig.~\ref{fig:3D_strings_and_EoS}.

\subsection{Results in two spacetime dimensions}
\label{subsec:2D}

In two spacetime dimensions (2D), several analytical (or semi-analytical) results have been known since the 1970's or early 1980's~\cite{tHooft:1974hx, Gross:1980he, Durhuus:1980nb}, but 2D theories continue to attract interest~\cite{Narayanan:2008he, Orland:2011rd, Orland:2012sk, Cubero:2012xi} as useful QCD toy models.

Recently, the eigenvalue density of Wilson loops in 2D has been investigated by Lohmayer, Neuberger and Wettig~\cite{Lohmayer:2009aw}; a related analysis in four dimensions has been carried out in ref.~\cite{Lohmayer:2011nq}.

Finally, there exist studies of 2D large-$N$ QCD at finite chemical potential~\cite{Bringoltz:2008iu, Bringoltz:2009ym, Galvez:2009rq}.

\section{Concluding remarks}
\label{sec:conclusions}

Lattice studies of gauge theories in the large-$N$ limit are theoretically very appealing, numerically tractable, and interesting for a very broad community. During the last fifteen years, numerical simulations in this field have given conclusive answers to various long-standing questions. However, many other issues are still open, and waiting for your involvement.

From my personal point of view, particularly promising research directions for further numerical studies at large $N$ include:
\begin{itemize}
\item simulations with dynamical fermions, in various representations, which are potentially interesting also for physics beyond the Standard Model;
\item further simulations at finite temperature and/or finite density, and comparisons with predictions from perturbative computations~\cite{Arnold:1994eb, Zhai:1995ac, Braaten:1995jr, Kajantie:2002wa, Andersen:2009tc}, from holography~\cite{Son:2007vk, CasalderreySolana:2011us}, or from phenomenological models;
\item studies of the topological properties: although some of the open problems in this context are numerically challenging, the results of previous works are encouraging~\cite{Lucini:2001rc, Cundy:2002hv, DelDebbio:2002xa, Lucini:2004yh, Vicari:2008jw}, and there is steady algorithmic progress~\cite{Panagopoulos:2011rb, DElia:2012vv};
\item further investigation of large-$N$ equivalences and volume reduction: in the past few years, this subfield has seen a revival of interest, with impressive progress both on the theoretical and on the numerical side, and further works are well motivated.
\end{itemize}

\vskip0.5cm \noindent {\bf Acknowledgements.}\\
The author warmly thanks the organizers of the conference for the invitation to present this review talk, and acknowledges financial support from the Academy of Finland (project 1134018).

\end{document}